\documentclass{article}

\usepackage{microtype}
\usepackage{graphicx}
\usepackage{booktabs}
\PassOptionsToPackage{hyphens}{url}
\PassOptionsToPackage{table}{xcolor}
\usepackage[breaklinks=true]{hyperref}
\usepackage{amsmath}
\usepackage{amssymb}
\usepackage{mathtools}
\usepackage{amsthm}
\usepackage[capitalize,noabbrev]{cleveref}
\usepackage{enumitem}
\usepackage[accepted]{icml2025}
\usepackage{booktabs}
\usepackage{subcaption}
\usepackage{tikz}
\usepackage{graphicx}  %
\usepackage{subcaption} %
\usepackage{float}   %
\usepackage[hang,flushmargin]{footmisc}

\makeatletter
\pgfdeclareverticalshading{agent_gradient}{100bp}{%
 color(0bp)=(black);
 color(50bp)=(gray!50);
 color(100bp)=(white)
}
\makeatother

\icmltitlerunning{Build Agent Advocates, Not Platform Agents}

\begin{document}

\twocolumn[
\icmltitle{Build Agent Advocates, Not Platform Agents}

\icmlsetsymbol{equal}{*}

\begin{icmlauthorlist}
\icmlauthor{Sayash Kapoor}{equal,princeton,mozilla}
\icmlauthor{Noam Kolt}{equal,hebrew}
\icmlauthor{Seth Lazar}{equal,anu}
\end{icmlauthorlist}

\icmlaffiliation{princeton}{Princeton University}
\icmlaffiliation{mozilla}{Mozilla}
\icmlaffiliation{hebrew}{Hebrew University}
\icmlaffiliation{anu}{Australian National University}

\icmlcorrespondingauthor{Sayash Kapoor}{sayashk@princeton.edu}
\icmlcorrespondingauthor{Noam Kolt}{noam.kolt@mail.huji.ac.il}
\icmlcorrespondingauthor{Seth Lazar}{seth.lazar@anu.edu.au}

\icmlkeywords{AI Agents, Platform Control, User Advocacy, Digital Rights}

\vskip 0.1in
]

\printAffiliationsAndNotice{\icmlEqualContribution} %

\begin{abstract}
Language model agents are poised to mediate how people navigate and act online. If the companies that already dominate internet search, communication, and commerce---or the firms trying to unseat them---control these agents, the resulting \emph{platform agents} will likely deepen surveillance, tighten lock-in, and further entrench incumbents. To resist that trajectory, this position paper argues that we should promote \emph{agent advocates}: user-controlled agents that safeguard individual autonomy and choice. Doing so demands three coordinated moves: broad public access to both compute and capable AI models that are not platform-owned, open interoperability and safety standards, and market regulation that prevents platforms from foreclosing competition.

\end{abstract}

\section{Introduction}
\begin{figure}[h!]
  \centering
  \includegraphics[width=\columnwidth]{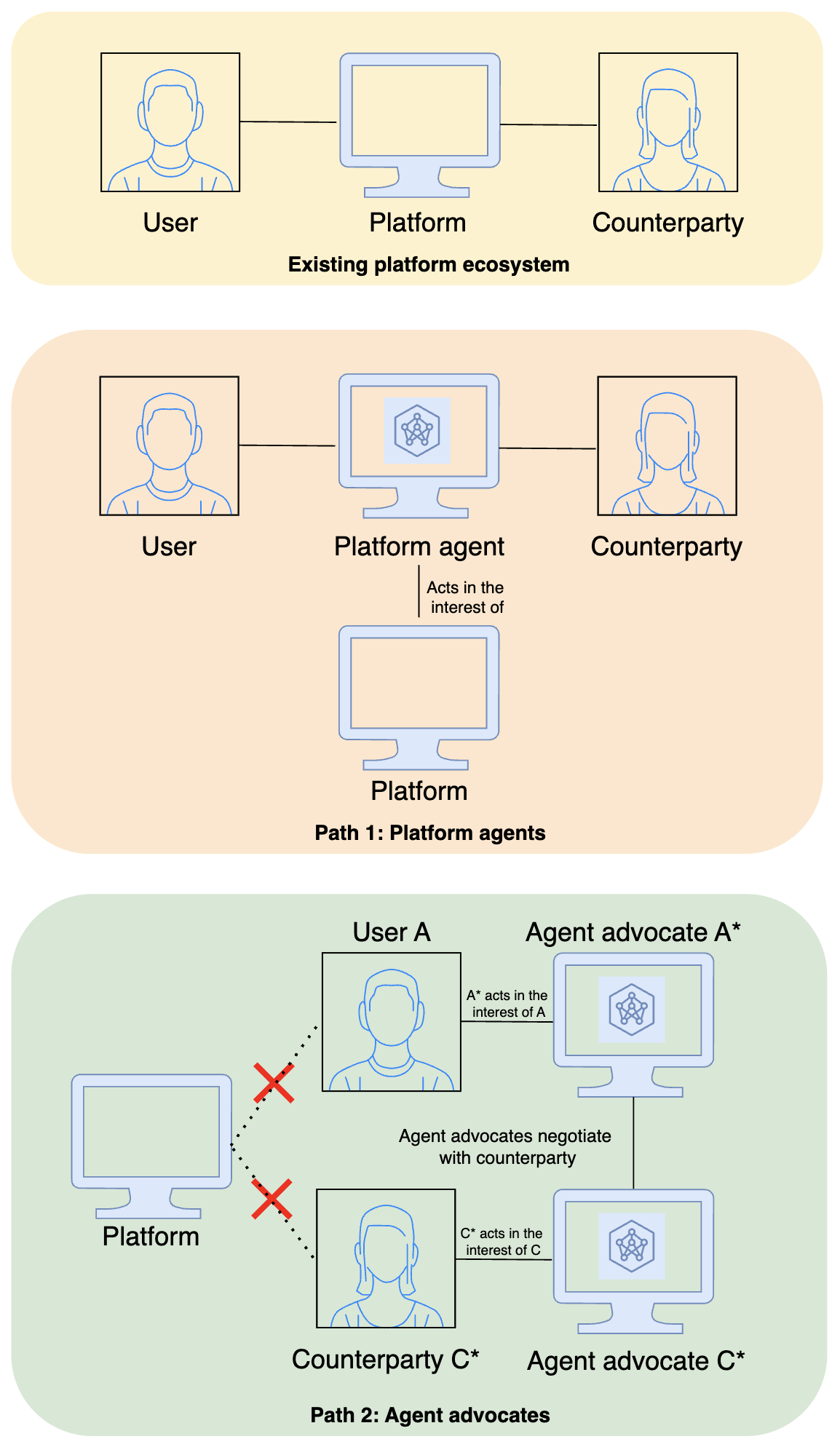} %
  \caption{Two development pathways for AI agents: \textit{platform agents} and \textit{agent advocates}. Platform agents intensify many of the concerns of the platform economy. Agent advocates that act in the interests of users (rather than platforms) could cut out the intermediary entirely.}
  \label{fig:agent-paths} %
  \vspace{-.75cm}
\end{figure}

Language Model Agents (LMAs) are compound AI systems that can plan and execute complex digital tasks with limited human input. They have become a major research and product focus of the major AI companies~\citep{wang2024survey, compound-ai-blog, casper2025ai}. Although current performance is still limited~\citep{updated_kapoor_ai_2025,hal}, significant progress has been made, especially with the introduction of reasoning models trained with reinforcement learning to undertake more complex tasks~\citep{openai_introducing_2024-1, google_deepmind_gemini_2025, anthropic_claude_2025, deepseek-ai_deepseek-r1_2025}. For example, Anthropic, Google DeepMind, and OpenAI have shipped LMAs that can use a web browser or virtual computer to perform a wide range of tasks~\citep{anthropic_introducing_2024,google_deepmind_project_2024,openai_introducing_2025-1}. Most AI companies now offer a `Deep Research' product that enables an LMA to conduct internet research on a topic and produce a corresponding report. OpenAI, Google, and a number of smaller companies have released software engineering agents that can undertake economically valuable coding work. Enterprise software companies are building agents intended to automate much of customer service~\citep{sierra_conversational_2025}. 

Platform companies---businesses that intermediate between users and products, services, and content---dominate the digital economy~\citep{RN3756, RN3668, Lazar2025, WuAgeExtraction}. Meta, Google, Amazon and Microsoft now aspire to develop the most prominent and capable LMAs, or else to steer other companies, such as Anthropic and OpenAI, that are leading the way in AI research~\citep{korinek2025concentrating}.\footnote{See, for example, \citet{wiggers_amazon_2024-1,wiggers_amazon_2024}.} Several major AI companies, such as OpenAI and xAI, now aspire to platform status themselves.\footnote{For OpenAI, see recent reporting such as \citet{updated_sullivan_openai_2025} and \citet{updated_temkin_openai_2025}; for xAI, see the merger of xAI with X (formerly Twitter), the social media platform.} Today's platform economy is likely to shape the emerging LMA ecosystem, for at least four reasons. \textit{First}, platform companies have a competitive edge in developing and shipping LMAs due to their unparalleled access to users and user data (including workflows on which to train their agents), as well as to create and deploy agent infrastructure that will cement first-mover advantage. \textit{Second}, users may themselves benefit from the convenience of using LMAs deployed by platforms with which they are already familiar. \textit{Third}, platform companies can acquire or otherwise impede new LMA companies that threaten their market position. \textit{Fourth}, LMAs offer a major opportunity to challenge platform incumbents' position, retaining the structure of the platform economy while changing those at its head.

By default, therefore, the first generation of successful LMAs will likely be \textit{platform agents}---controlled by incumbent or aspirant platform companies, acting on their behalf, and following their business logic. In particular, platform agents are likely to amplify the power that platform companies already hold, by: enabling aggravated surveillance harms; more decisively controlling the allocation of user attention; extracting additional rent from transactions that they intermediate; and more assertively policing user behavior.

While platform agents are the default path, this outcome is \textit{not} inevitable. LMAs could genuinely advance user interests, not platforms'. In contrast to platform agents, \textit{agent advocates} would serve user interests without competing obligations to platforms. They could run not on the platform's servers, but on local hardware or an encrypted private cloud, giving users visibility into and control over their actions and data collection. \textbf{In this paper, we argue that AI researchers and developers should resist the rise of platform agents and spur the development of user-centric agent advocates.} 

We first describe platform power in the broader digital economy (Section 2), then discuss the risks of platform agents as the default path of LMA development (Section 3). In Section 4 we show that agent advocates can mitigate platform agents' harms, and explore how technical and institutional challenges to their development can be addressed. In Section 5, we consider objections, and in Section 6 discuss complementary proposals for resisting platform agents. We conclude by advocating for three interventions to support agent advocates: broad public access to compute and independent AI models, open interoperability and technical safety standards, and market regulation that prevents platforms foreclosing competition.

\section{Background: Platform Power}
\vspace{-.25\baselineskip}

Digital platforms are intermediaries that enable interactions and exchanges among users and between users and businesses~\citep{RN3756, RN3928, Lazar2025}. For example, Amazon mediates e-commerce, Google mediates search, and Meta mediates online communication. In exchange, platforms take a `cut' from the interactions that they mediate---often by collecting personal data that provides insights into user behavior~\citep{kolt2019return} or capturing user attention for advertising~\citep{RN2932, hwang_subprime_2020, RN3714}.

Intermediaries mediate between two or more \textit{principals}. Schematically, intermediaries vary along three graduated dimensions. The first concerns whether the intermediary is a \textit{representative} or a \textit{go-between}. A representative acts on behalf of one principal. A go-between goes back and forth between the principals. For example, a lawyer should be a representative, while matchmakers and more conniving intermediaries like Iago in Shakespeare's \textit{Othello}, are go-betweens. The second dimension concerns whether the intermediary advances the principal's interests, or someone else's (its own or a third party's). How much, if any, value does the intermediary extract from the transaction that it mediates? The third dimension concerns the degree to which the intermediary \textit{constitutes} the relationships it mediates, by determining what is possible or impossible within it. \textit{Neutral} intermediaries are content-independent pipelines; \textit{constitutive} intermediaries substantially shape the relationships that they intermediate~\citep{Lazar2025}.

Current digital platforms are \textit{constitutive}, \textit{self-interested} \textit{go-betweens} that extract value from the principals they mediate between~\citep{RN3756, RN3928, RN3564}. We highlight four key trends: (a) platforms subject users to pervasive surveillance; (b) they establish pseudo-markets that depart from liberal ideals; (c) they substantially determine the allocation of user attention; and (d) they define and enforce policies that shape user behavior. 

Ironically, early warnings about the surveillance practices of digital platforms were so hyperbolic and regulatory interventions so ineffective (e.g., cookie banners), that digital surveillance now inspires widespread complacency. But, as a recent FTC report reconfirmed, we are all still exhaustively tracked across the internet, in ways that can readily be reconstructed for advertising or other purposes, such as identifying possible targets for immigration enforcement.\footnote{See \citet{federal_trade_commission_ftc_2024}; this confirmed years of research. For a comprehensive overview of the relevant literature see \citet{RN3714}. For recent reporting on the use of AI for immigration enforcement, see \citet{haskins_ice_2025}.} 

Digital platforms initially emerged as two-sided market-makers that mediate between those with specific needs and others who could fulfill them~\citep{RN3753,RN3678}. This function was critical in the mid 2000s, as rapid internet growth introduced significant challenges around filtering, matching, and trusting online interactions~\citep{RN4073}. However, rather than facilitate genuine markets---whether for goods or ideas---in which participants freely choose among impartially presented counterparties, platforms increasingly manipulated these exchanges to extract value from both sides of these transactions~\citep{RN3569,RN3564}. In Cory Doctorow’s terms, platforms have “enshittified”~\citep{doctorow_enshittification_2023}. This goes beyond simply charging a commission or harvesting transactional data such as price signals and consumer behavior patterns. Platforms also shape incentives to advantage favored actors---including themselves~\citep{farronato2023self}---and actively obstruct interactions that might otherwise occur outside their domain (for example, when eBay identifies and prohibits attempts to share contact information with another user through their platform)~\citep{RN3569}.

Platform companies also shape the distribution of attention. For example, Google dominates online search, with close to 90\% of market share ~\citep{statcounter_search_2024}. Meta's Facebook, Instagram, and WhatsApp account for three of the top four social media platforms, with over 3 billion users (YouTube is ranked \#2)~\citep{shopify_most_2024}.\footnote{While digital platforms are prevalent worldwide, we focus on their impact in the U.S., primarily because that is the context we are most familiar with, and where major platforms are based.} By shaping how individual users' attention is allocated, platforms influence how people form their beliefs and desires~\citep{Susser2019, RN3714}. In addition, their impact on the allocation of collective attention profoundly influences societies' capacity and appetite for collective action~\citep{RN5479}.

Digital platforms also exercise significant \textit{governing} power over users---they make, implement, and enforce rules that users are compelled to abide by~\citep{Lazar2025, Black2002}. Researchers have long argued that these rules---for example, about which content may be posted online, or how such content will be amplified or `de-boosted'---are substantively unjustified and procedurally illegitimate~\citep{Gillespie2018, suzor2018}. In recent years, this concern has spread as users and major political parties have expressed and acted on concerns about illegitimate governance of online speech~\citep{nunziato2023old}. 

These platform companies are now racing to build capable LMAs. Younger, AI-first companies such as OpenAI and xAI aspire to \textit{become }platforms---to take market share in search, e-commerce, social media etc. from the established companies. The first highly capable LMAs are likely to be shaped by the structures of the existing platform economy.

\section{The Default Path: Platform Agents}
\vspace{-.25\baselineskip}

\textbf{Defining and scoping language model agents (LMAs).} An LMA is an \textit{agent} insofar as it can plan and act (agent is derived from the Latin for ``to do") to achieve goals in complex environments with limited human input~\citep{chan_harms_2023, shavit_practices_2023, lazar2024frontier, gabriel_ethics_2024, updated_kapoor_ai_2025, kolt2025governing}.
An LMA is a \textit{language model} agent insofar as calls to one or more LLMs play a crucial role in its architecture, typically scaffolded with tools that enable it to perceive, plan, and act in its environment~\citep{updated_sumers_cognitive_2024, casper2025ai}.

Initially, LMAs will operate in narrow, task-specific domains---much like current `Deep Research' agents, though in more domains. For instance, one agent might monitor and reply to work emails, another might order groceries, yet another might keep track of financial markets, perhaps making investments on a user's behalf. At least from the user's perspective, these narrow-purpose agents will ultimately be subsumed into a single digital concierge that mediates between users and most or all of the digital environments with which they interact. These \textit{universal intermediaries}~\citep{lazar2024frontier} will be able to use any app to perform tasks on behalf of users, not only to retrieve information to inform user action. In doing so, they will draw on their model of the user derived from all of their different interactions, to offer a more personalized and useful service. We expect LMAs to ultimately undertake most tasks for which people currently use the internet---including the procurement of products, services, and content.

\textbf{What are platform agents?} Platform agents are LMAs molded by the incentives and selection pressures of the platform economy to function as constitutive go-betweens controlled by incumbent or aspiring platform companies, and designed to act in the platform companies' interests. 

\subsection{Risks from platform agents} 
\vspace{-.25\baselineskip}

Today's algorithmic intermediaries are already self-interested, constitutive go-betweens. Platform agents, due both to their ability to take \textit{actions}, and their capacity for developing a much richer and more nuanced model of the user, would inherit and intensify the aspects of platform power described above.
 
\textbf{Heightened surveillance harms.} Platform surveillance harms users primarily by enabling interventions on their behavior that are against their own interests. Existing recommenders predict user behavior at scale, making statistical inferences based on behavioral data collected from billions of users. This has been described as a kind of \textit{stochastic manipulation}~\citep{RN3714}: though effective at the margin when applied to massive populations, it is usually harmless at the individual level (see, e.g., very low click-through rates on targeted advertisements ~\citep{hwang_subprime_2020}). Platform \textit{agents} will be able to make much more individualized interventions, by drawing on their detailed, even intimate knowledge of their users across many distinct interactions. They won't rely on statistical patterns alone, but will know their users like a close friend. This knowledge, deployed by a self-interested commercial entity, may lead to much more acute manipulation harms than are now feasible. 

\textbf{Market design.} Platform companies' current attempts to shape quasi-markets in their favor can sometimes be resisted. For example, customers and vendors can occasionally cut out the intermediary and transact directly (e.g., by arranging a second visit to an Airbnb directly with the owner). Similarly, users who invest time in independently researching products, services, or content can escape platforms' walled gardens, albeit paying the cost of inconvenience. As LMA capabilities improve and users rely on platform agents to conduct these everyday transactions, they will have progressively less opportunity to conduct their own basic due diligence, giving platform companies even greater control over their market choices---most obviously by favoring their own products, or those of vendors who buy advertisements~\citep{jeffries_amazon_2021, updated_fowler_its_2022} or strike revenue-sharing deals with the platform.

Relatedly, platform agents could exacerbate the already acute switching costs and network effects of existing platforms, by designing systems to work most effectively with agents from the same platform---as forewarned by some prescient legal scholars~\citep{gal2016algorithmic, van2018digital}. Switching away from a general-purpose digital concierge will be even more difficult than exiting a standalone platform website, because LMAs will know users intimately, learning about them over the course of perhaps years' worth of interactions. This tacit knowledge will afford a vastly better service, locking users in. This is already in train, with OpenAI enabling ChatGPT to reference all past conversations, and Google's Gemini being able to integrate a user's search history. 

\enlargethispage{\baselineskip}
\textbf{User attention and engagement.} Platform agents could more minutely influence users' attention, compared with existing platform technologies. Despite prevalent concerns about filter bubbles and echo chambers in social media \citep{RN3958, RN2824}, users can still piece together their media consumption across different platforms and are invariably exposed to some diversity of content~\citep{RN3958, RN2824, bruns2019filter, cinelli2021echo, sharma2024}. If platform agents browse the open internet on users' behalf, they will not only control users' information diet, but also how it is presented to them---enabling pervasive editorializing that would be very likely to reshape users' beliefs and desires over time.\footnote{This is already apparent even just with opinionated autocorrect models~\citep{jakesch2023co}.} For example, platform agents could subtly nudge users through dialogue, or manipulate users by expressing approval or disapproval of the users' tastes, decisions, and behaviors~\citep{RN4721}. Just as platform agents can influence market supply, they can also influence market demand.

\textbf{Intensified governance by platform companies.} Platform agents could prove invaluable tools for illegitimately governing users~\citep{Gillespie2018, suzor2018}. By designing guardrails and nudges into LMAs, much as they are now built into LLMs, fine-grained control over people's use of digital technologies will become possible. The more pervasive a platform agent's intermediary role, the greater its ability to monitor our digital actions and interactions for conformity to some set of behavioral standards, as well as its ability to preclude or preempt undesired action \citep{RN4117}. ``Alignment" techniques can, in other words, be used to limit users' freedom of action. While these techniques are not at present robust to adversarial attack, they are sufficiently constraining for most ordinary users that they can empower platform companies to establish and preemptively enforce compliance with platform rules---irrespective of those rules' legitimacy~\citep{RN4615}.

\enlargethispage{\baselineskip}

\subsection{Platform agents are the default path.} 
\vspace{-.25\baselineskip}
Platform agents will likely increase the \textit{concentration}, \textit{degree}, and \textit{scope} of platform power. As platform agents become more capable, they will be and used by more people while still being controllable by platform companies and ultimately by those companies' CEOs (concentration). They will be used in more consequential interactions (degree), and across a wider range of actions (scope). We see this as the default pathway for LMAs, for the following reasons:

\textbf{Path dependency.} New technologies are likely to be molded by the political economy from which they emerge. Platform companies dominate the digital economy. They have clear incentives to capture the emerging market for LMAs---or else risk losing that dominance.

\textbf{User base.} Platform companies are incumbents with built-in user bases. They do not have to compete for an initial audience for LMAs. They can also use their existing user base to obtain feedback, as well as workflow data, that allows building more effective agents. 

\textbf{Realization of AI investment.} Platform companies have already invested billions in training AI models. They will expect returns on that investment. Platform agents are an obvious way to realize those returns~\citep{stripe_adding_2024,linkedin_meet_2024}. Some major platforms may even develop platform agents if only to prevent competitors' agents from drawing user revenue and attention away from them.

\textbf{Competition and incumbents.} The platform economy favors a small number of large companies. The advent of the agent economy presents an historically unprecedented opportunity to unseat at least some incumbents. This is a powerful incentive for other companies to aim at the near-hegemonic status of existing platform companies. In response, incumbents will likely take steps to defend their market position, acquiring new contenders such as smaller LMA developers where possible, undercutting them and undermining them where not \citep{updated_mehta_exclusive_2024,wiggers_amazon_2024-1}.

\section{Our Proposal: Agent Advocates}
\vspace{-.25\baselineskip}

\textbf{What are agent advocates?} Platform agents are not the only path for LMA development. Instead of agents being a self-interested constitutive go-between controlled by and loyal to platform companies, they could be designed to represent the users who operate them, and act exclusively in their interests. Pursuing this \textit{agent advocates} path would not only forestall platform agents' worst harms, but would also provide users with tools to shift the balance of power in the digital economy away from platforms.\footnote{We note here complementary work on fiduciary AI by~\citet{benthall2023designing, updated_reisman_new_2024}.}

\subsection{Addressing the risks of platform agents.}
\vspace{-.25\baselineskip}

\textbf{Decreasing surveillance harms.} At a minimum, agents that advance the user's and not the platform's interests would avoid the surveillance harms that platform agents would cause. They could give users meaningful control over their data, either operating locally or in the private cloud. Beyond this, agent advocates could actively \textit{defend} users against platform surveillance. For example, if agent advocates can browse digital platforms on the user's behalf, extracting information to re-present to them, they can deliver those platforms' benefits without the downsides of surveillance and targeted advertising~\citep{updated_lazar_moral_2024}. They could also be designed to detect and interdict any attempts by platforms (or others) to track them online.

\textbf{Confronting platforms' market design power.} Platforms' market design power derives from their being a go-between that acts for both sides of the transactions that they mediate. If each party to online transactions has their own representative acting in their interests, the scope for this kind of confected quasi-market is reduced. Instead of communicating only through a centralized and self-interested intermediary, parties in two-sided markets could communicate with one another through representatives acting in their respective interests, enabling new approaches to market design. Agent advocates would still depend on \textit{some} matchmaking service, but they would be better-positioned than human users to monitor those matchmakers for predatory activity and to switch to better alternatives, as well as to communicate directly with counterparties. More than just avoiding platform agents' worst outcomes, agent advocates could recalibrate the digital economy towards more competitive markets.

\textbf{Reducing platform lock-in.} Current platforms bundle together several different functions~\citep{RN4073, Gillespie2018} and create walled gardens around those functions, relying on switching costs and network effects for user retention. Companies that offer competing services (such as DuckDuckGo for search, or Brave as a browser) are at a systematic disadvantage, simply because exiting the platform ecosystem requires more effort on the user's part, and potentially shuts one out from the benefits of joint usage of the same service by others (this is especially true for messaging services). Agent advocates could reduce switching costs by reducing friction in using new services, and could reduce network effects by providing a kind of bottom-up interoperability, hopping over the platforms' walled gardens. Consider, for example, an agent advocate that can collate various social media feeds, or create a group chat that integrates users of different messaging services. 

Platforms fragment users, undermining collective action and inducing `digital resignation'~\citep{RN2755, RN3714}. Many have advocated for collective action by users to resist this trend---but it does not happen because it is too effortful. Agent advocates could undertake some of the effort of coordinating on users' behalf.

\textbf{Changing how user attention translates to revenue.} Platforms determine how content creators are paid and shape how attention is allocated. Agent advocates could enable new approaches to matching attention to revenue. For example, users currently pay for content through subscriptions and a hodgepodge of free trials and other kludges. Users do this instead of paying for each individual piece of content in part because the transaction cost of deciding whether to outlay some small sum for some content is greater than the value of the content itself. Users could outsource this cognitive effort to their agent advocate, to decide whether the content is worth paying for, at what price. LMAs would also be able to provably forget content, so a user's agent can `view' something on the user's behalf, and if it is not worth paying for, can provably forget it so that the publisher has lost nothing~\citep{updated_weiss_redesigning_2024}. More generally, the owners of large social media websites currently have disproportionate control over the allocation of collective attention; agent advocates could enable users to decide for themselves what to attend to~\citep{RN5479, updated_lazar_moral_2024}. This could of course have some negative implications, but still valuably shifts power away from platforms. 

\textbf{Increased user autonomy.} Platforms today can unilaterally shape how users interact and extract as much value from that interaction as users will tolerate~\citep{WuAgeExtraction}, while subjecting them to often dubious and sometimes clearly illegitimate governance. Users endure this unfair bargain because they have no reasonable alternative, and because for all their predations, platforms do offer significant value. Because of this, and because platforms' rules are, for most users, neither onerous to follow nor likely to result in highly costly penalties, most do not experience their quasi-forced subjection to platform governance as an acute kind of unfreedom. However, platforms' growing role in users' lives, and the prospect that platform agents might become a pervasive intermediary to \textit{all} of our digital actions and interactions, suggests that this will change. Agent advocates are a bottom-up response that work not only to extend users' agency by acting as their representative, but to actively combat platform power. Two parties who can only interact through a self-interested go-between, and who depend on that interaction to some non-trivial degree, are subject to the arbitrary power of that go-between. Two parties who can instead interact through fiduciary AI representatives are not subject to the same arbitrary power.

\subsection{Interventions to facilitate the development and adoption of agent advocates}
\vspace{-.25\baselineskip}

Platform agents and agent advocates present two extreme points on a continuum. The default path skews towards the first extreme, where LMAs function as self-interested go-betweens that extract value from users and exercise power over them. How can we shift the default trajectory towards genuinely user-centric agent advocates that act as faithful representatives of their principal?

\textbf{Availability of useful open models.} Small AI companies with no prospect of unseating the platform incumbents might be able to build effective LMAs by integrating calls to closed frontier language models. Companies like OpenAI and Anthropic \textit{could} in this way support the development of agent advocates, instead of pursuing the platform agent path. But that is too fragile a foundation for the future of the digital economy. Open weights models are likely essential to reduce exposure to the risks of a reversal in corporate strategy, and to give users real guarantees that an agent advocate is acting only in their interests ~\citep{updated_kapoor_position_2024} (technical barriers are discussed in Section 5.2).\footnote{Of course, we assume that few users will build their own agents; but the availability of capable open models will allow small companies to develop user-centric agent advocates.}

Currently, platform companies like Meta and Google and Chinese companies DeepSeek and Alibaba (itself a platform company) lead the way in open AI development~\citep{RN5180,deepseek-ai_deepseek-r1_2025,team_qwen25-max_2025}. Mistral~\citep{mistral_ai_team_cheaper_2024} and Cohere~\citep{cohere_coheres_2024} also offer relatively competitive open models; AI2, a nonprofit research institute, though behind in terms of capabilities, has shown the returns on making models, code, and data entirely open~\citep{updated_groeneveld_olmo_2024}. The open foundation model ecosystem is today in better health than ever before, and yet access to the most capable models still depends on the whims of Silicon Valley executives and, for Chinese models, ultimately the leadership of the CCP. This is an unstable position for agent advocates. Meta and Google, for example, are unlikely to look favorably on LMAs that use their models to draw revenue away from their platforms---indeed, these will likely be prohibited by the license under which these models are provided. Chinese models, meanwhile, will have to comply with rules promulgated by the CCP. At present this kind of censorship can be circumvented~\citep{perplexity_r1_1776}, but that may not always be so. 

Expanding the availability of 
open models is an urgent imperative. Liberal democracies should create a public complement to the private industry leaders in frontier AI research. Researchers advocating a `CERN for AI' often aspire primarily to help states better understand the risks of advanced AI, and unlock the societal benefits of `superintelligent' AI systems~\citep{brundage_my_2024}. While these are important initiatives, public AI could also be leveraged to prevent corporate or political control over advanced AI agents that would otherwise have severe negative implications for individual and collective autonomy and well-being~\citep{public_ai_network_public_2024,marda_creating_2024}. 

In addition, while we might hope that one day agent advocates might be powered by open models running on device, this is likely to be technically infeasible for some time. And if the computational demands of effective agent advocates continue to exceed what can feasibly be deployed locally, then the provision of affordable computational power (compute) will be as important as the availability of public interest AI models. 

\enlargethispage{\baselineskip}

\textbf{Bolstering competitive markets.} Agent advocates will threaten platform companies' market dominance. Those companies will restrict agent advocates however they can. Competition regulators will need to intervene to prevent this. For example, they could disbar digital platforms from blocking AI agents from their GUIs, and perhaps even require them to make APIs available for appropriately authorized agents to access. This intervention could be motivated on similar grounds to the `net neutrality' principle~\citep{wu2003network} or earlier attempts to enforce interoperability~\citep{wegner1996}. This would enhance competition by giving users a real threat of exit, and by changing how advertising revenue is distributed. It would also incentivize competitors to produce better AI models or make better use of existing models to have a chance at the new value created by LMAs. Similarly, governments should explore the application of existing regulations that protect data portability to LMAs from the outset, ensuring that users are able to switch away from a given provider whenever they wish to.

\textbf{Operational and regulatory infrastructure.} By centralizing control, platforms make effective governance of users comparatively feasible; indeed, Tarleton Gillespie has argued that effective governance is the central product that platforms provide~\citep{Gillespie2018}. A decentralized economy of agent advocates would be more liberal than a platform agent economy. But it is also riskier. For example, how can a user trust that their advocate is robustly advancing their interests? When agents transact with one another, what assurance do users have that they are not colluding, or that one has not manipulated the other? A decentralized economy of agent advocates will require novel infrastructure to ensure it is safe and fair~\citep{updated_chan_infrastructure_2025}. 
 
You cannot govern what you cannot identify. Agent advocates will require infrastructure to provide agent instances with \textbf{credentials} ~\citep{updated_chan_ids_2024, adler2024personhood, chaffer_decentralized_2025,south_authenticated_2025} that certify that they meet certain minimum criteria to operate autonomously online. These credentials should also provide key information such as the agent's underlying architecture, its core capabilities, the tools to which it has access, the types of actions it can take, and some identifier (likely anonymized) that attributes the agent's actions to some upstream individual or company who can be held to account for its actions. They could also include credentials that certify the agent's adherence to a fiduciary responsibility to the principal.

Platforms create network effects. To build an agent economy without platforms, there must be protocols that enable interoperability~\citep{RN4231}. These would include \textbf{communications protocols} for agents (even though they can communicate in natural language this is unlikely to be maximally efficient ~\citep{RN5691}). Open communication protocols can prevent the establishment of walled gardens by platform companies. But they can also be used as mechanisms for governance, to incentivize credentialing. For example, users would be able to authorize their agents only to communicate with other credentialed agents. 

An internet of agent advocates will be a vast market constituted by billions of microtransactions. Because directly regulating these microtransactions will be impractical, digital \textbf{clearinghouses} will be needed to secure agent transactions, and both protect the principals' interests against collusion and fraud, and monitor for and mitigate negative externalities. Clearinghouses will ensure that agents do not manipulate or prompt inject each other, and provide means for recourse in the event of misconduct. This function could potentially be incorporated into communication protocols, but more likely would need to be routed through independent, secure servers, on a similar model to the existing payments infrastructure for online transactions. Clearinghouses would provide further opportunity for bottom-up, market-based governance of LMAs, as agent credentials could be a necessary condition for transacting, and users could specify that their agents are not permitted to transact with other agents with specific properties---for example, you might instruct your agent to transact only with other agents of the same capability level (this might help reduce the risk of an arms-race dynamic in the open agent economy).\footnote{Would clearinghouses potentially become the new platforms? We don't think so. Platforms are defined by their bundling of different functions; clearinghouses would play a discrete role of providing bottom-up market-based governance.}

As the development of these protocols and clearinghouses progresses, their implementation and promotion should be supported by \textbf{technical standards}. These standards, which would be comparable to existing standards for networking and cybersecurity~\citep{ietf_internet_2024}, could be promulgated either by public bodies, such as the National Institute of Standards and Technology (NIST), or private bodies, such as the International Organization for Standardization (ISO). 

The successful adoption of agent advocates requires \textbf{mitigating the risk of misuse and misalignment}. While developing robust credentials and protocols would facilitate the creation of agent advocates, they will not address the risk of malicious actors using agent advocates for nefarious purposes, or agents that malfunction or are misaligned with the interests of users or broader societal values. Mechanisms will need to be established to detect and disrupt agent advocates in such cases, which could potentially be achieved by developing defensive agents that identify and intercept malign agents.

\section{Alternative Views}
\vspace{-.25\baselineskip}

\subsection{Counterargument 1: Platform agents are not a problem.} 
\vspace{-.25\baselineskip}

\textbf{Development and deployment bottlenecks.}
Despite growing research and investment in LMAs, these systems may not become sufficiently capable or reliable for large-scale adoption, whether by platforms or other actors. Key bottlenecks in LMA development and deployment include: (i) the exceptionally high levels of reliability needed to perform complex sequential actions with only limited human involvement; (ii) the ability to generalize robustly and act effectively in diverse digital environments; (iii) the need to collect or create training data comprised of `agentic trajectories' (i.e., state-action sequences); (iv) expensive operating costs \cite{updated_kapoor_ai_2025, hal}.

\textbf{Market responses.}
Even if LMAs become highly performant and sufficiently cheap, the risks considered above may not materialize if other market actors (e.g., online vendors, news sites) obstruct platform agents. Such measures may be technical (e.g., protocols resembling robots.txt and LLMs.txt adapted to agents specifically) or legal (e.g., access and licensing agreements with platform companies). Platform companies may also refrain from deploying LMAs as a B2C product that intermediates between consumers and digital services and, instead, integrate LMAs into their internal business operations, or use them to support external B2B services (e.g., managing platform-vendor relationships).

\textbf{Advantages of platform agents.}
Platform agents may offer advantages over agent advocates that counterbalance their risks. For example, platform agents are likely to be comparatively convenient, due to the combination of data, compute, engineering talent, and safety and security practices, available to platform companies. Accordingly, just as hundreds of millions of consumers currently use platform-operated chatbots~\citep{backlinko2025chatgpt} despite the availability of non-platform alternatives, they may also prefer to use platform-controlled AI agents over agent advocates. 

\textit{\textbf{Our response. }Even if platform agents do not develop precisely as discussed in Section 3, the prospect of even some platform-based risks arising from LMAs should nevertheless motivate the promotion of agent advocates. There is clearly a spectrum between platform agents and agent advocates; we argue for moving as far along that spectrum towards advocates as is feasible, subject to appropriate safety standards.}

\subsection{Counterargument 2: Agent advocates are not feasible.} 
\vspace{-.25\baselineskip}

\textbf{Technical and safety barriers.}
While researchers have explored the notions of `loyal AI' and `fiduciary AI'~\citep{aguirre2020ai, benthall2023designing} and built systems that follow user instructions \cite{ouyang2022training, zhou_language_2023}, the problem of ensuring an agent's loyalty to its principal is far from solved. The AI alignment literature~\citep{russell2019human, christian2020alignment, ngo2022alignment} illustrates myriad ways in which agents may fail to comply with user instructions, or comply in undesirable ways. Agent advocates are not immune to these problems, and we need to expand ways to control agents' actions. For these and other reasons, some researchers advocate against building autonomous agents altogether~\citep{cohen2024regulating, mitchell2025fully, bengio2025superintelligent}.

\enlargethispage{\baselineskip}

\textbf{Institutional concerns.} 
We argue that platform companies' profit incentive will drive them to develop self-advantaging LMAs that, at times, harm consumers; it will also drive other AI companies to aspire to build platform agents instead of agent advocates. The same incentives will also influence companies that develop and deploy agent advocates. Even if those companies initially develop user-centric agents, they may decide to change direction in the future, perhaps building platform agents of their own. 

\textbf{Past failures.}
Prior attempts to challenge digital platforms and decentralize the control of digital technologies have often failed~\citep{narayanan2012critical}. Efforts to develop agent advocates might meet a similar fate.

\textit{\textbf{Our response. }Agent advocates face serious obstacles. For this reason we propose the interventions outlined in Section 4.2. In addition, LMAs present more favorable conditions for mitigating platform power than today's internet infrastructure, since (a) They can both escape network effects and actively undermine them; and (b) User-centric companies can unilaterally build LMAs that, with sufficiently capable models, will undermine the power of platform companies.}

\subsection{Counterargument 3: Agent advocates fail to address important problems.} 
\vspace{-.25\baselineskip}

\textbf{Malicious applications.}
LMAs can be used for various malicious applications, including fraud, cybercrime, harassment, and weapons development \cite{updated_andriushchenko_agentharm_2025, updated_li_wmdp_2024, tur2025safearena}. Despite attempts to curtail such applications, jailbreaking is widespread and appears likely to persist~\citep{updated_wei_jailbroken_2023, updated_xu_comprehensive_2024}. Developing agent advocates does not address this pressing issue, and may even embolden malicious actors.

\textbf{Multi-agent interactions.} 
LMAs' most significant challenges may arise not from individual user-system interactions, but from multi-party interactions between different AI systems, human actors, institutions, and environments~\citep{dafoe2020open, updated_dafoe_cooperative_2021, hammond2025}. These include collusion, collective action problems, emergent network effects, and correlated or cascading failures \cite{updated_motwani_secret_2024, updated_anwar_foundational_2024, kolt2025complexl}. While user-centric agent advocates do not necessarily exacerbate these concerns, they do not address them either.

\textbf{Broader societal impacts.}
LMAs could have significant impacts on social and economic life. For example, interpersonal encounters might be increasingly mediated or even altogether obviated by LMAs~\citep{lazar2024frontier, kolt2025governing}. Certain jobs or workplace tasks might be outsourced to LMAs, either displacing people or altering the kind of work they undertake ~\citep{sellen2024rise}. 

\enlargethispage{\baselineskip}

\textit{\textbf{Our response. }Agent advocates will not solve all the problems of LMAs; they specifically target harms from platform agents. Supported by appropriate safety and governance frameworks, agent advocates are a pareto improvement on the default trajectory described in Section 3.}

\section{Expanding the Solution Space}
\vspace{-.25\baselineskip}

Agent advocates are not the only way to address risks from platform agents. Other approaches can either complement agent advocates or operate effectively even in their absence. We focus on proposals that aim to preempt such harms, rather than redress them after the fact. 

\textbf{Comprehensive AI regulation.}
One potential approach to mitigating platform agent risks is to pass comprehensive AI regulation that prohibits or severely curtails their most harmful consequences. For example, bans could be placed on the most intrusive surveillance practices and forms of user manipulation (discussed in Section 3 above). Prohibitions like these have already been established in the European Union in Article 5 of the EU AI Act~\citep{euro2024}. 
\textit{While such provisions could in theory address certain risks from platform agents, the prospect of the U.S. passing federal legislation comparable to the EU AI Act is highly unlikely.}

\textbf{Platform regulation.}
Regulation could instead target platform companies, rather than AI technologies more generally. For example, large platforms could be prohibited from controlling agents that intermediate between users and particular services, perhaps in sensitive settings such as banking and healthcare. Alternatively, platforms might be subjected to rigorous transparency requirements and external audits.
\textit{While the EU's Digital Services Act~\citep{DSA2022} offers a model for such regulation, the U.S. has largely refrained from regulating platform companies (e.g., social media), and appears unlikely to change course. 
}

\textbf{Consumer privacy law.}
LMAs will both rely on and generate large quantities of private data. Data protection regulation could offer an effective, if indirect, strategy to address risks from platform agents~\citep{gupta2024data}. Imposing strict controls on the uses of personal data for training or operating LMAs or requiring real-time informed consent from users could curtail certain undesirable consequences. 
While there are currently no broadly applicable federal consumer privacy laws in the U.S., some large states, including California, have passed noteworthy consumer privacy laws~\citep{CCPA_2018}.
\textit{Whether such laws could be amended or otherwise tailored to tackle the distinct challenges posed by LMAs is, for now, an open question.}

\textbf{Antitrust law and enforcement.}
In the absence of new regulation, existing antitrust law could offer a strategy for addressing platform agent risks~\citep{narechania2024antimonopoly}. Companies that undertake some combination of developing chips, providing cloud computing, training frontier models, building products (including LMAs), and operating consumer and enterprise platforms have a substantial advantage over companies that operate at only one point in the value chain~\cite{korinek2025concentrating}. Constraining such vertical integration could significantly bolster competition in the LMA industry, allowing new entrants to better compete.
\textit{A key obstacle to this approach is that the U.S. (in contrast to the EU) does not have a history of successfully taking major antitrust enforcement against platform companies or digital services providers more broadly, although this may yet change.
}
\section{Conclusion}
\vspace{-.25\baselineskip}

Computing amplifies agency. In the hands of the powerful, it reinforces centralized control. In individuals' hands, it can enable counter-power. Historically, recurrent moments of technological expansion have seemed poised to usher in a more decentralized computing future. Each time, however, centralizing forces have reasserted themselves. Examples abound: the first hackers circumventing the gatekeepers of MIT's PDP-6; the Silicon Valley Homebrew Club building alternatives to IBM's mainframes; open, customizable software vs. closed operating systems; community-run BBSs vs. centralized ISPs; the open internet standing against the internet of platforms, and more. 

Our current moment is not unique. It may, however, present a unique opportunity. Previously, the pathway toward decentralization was accessible primarily to technologically skilled users---hackers capable of circumventing constraints set by centralized authorities. Today, however, user-centric agent advocates could level the playing field.

By default, AI agents are likely to follow the same centralized pattern as the platform economy. Incumbent and aspiring platform companies will develop and control powerful agentic systems. These \textit{platform agents} will intermediate digital interactions across countless personal and professional contexts. Although users may guide platform agents, ultimate control will remain firmly with centralized developers. Platform-controlled AI agents will be double agents, with the potential for profoundly negative implications: heightened surveillance harms, constrained user choice, granular market manipulation, and broad illegitimate power. The worst of platform capitalism's current ills could be exacerbated.

But this outcome is not inevitable. A compelling alternative exists: user-centric \textit{agent advocates} designed to serve the interests of individual users, not platform companies. Representatives, not go-betweens, that reject platform logic. Agent advocates could provide a path to harnessing the promise of AI agents without succumbing to platform-based control. Realizing this decentralized alternative will require targeted technical and institutional interventions. These include ensuring the availability of open-source models and public computational resources, as well as establishing robust safety standards and governance frameworks. It will also require engineers who can build highly capable universal intermediaries but resist entering the race to create the next platform. Independent researchers and developers must prioritize addressing these challenges now---before the default pathway locks in.

\section*{Acknowledgments} 

We are grateful to Markus Anderljung, Matt Boulos, Alan Chan, Simon Goldstein, Lewis Hammond, Arvind Narayanan, Jake Stone, and members of the MINT lab at the Australian National University for feedback on a draft of this paper. SL acknowledges funding from the Australian Research Council (Grant FT210100724) and the Templeton World Charity Foundation.

\bibliography{references,references-1,references-2,references-3,reference-updates}
\bibliographystyle{icml2025}

\begin{table*}[t]
\centering
\vspace{0.5cm}
\section*{Appendix: Hypothetical Examples of Platform Agents and Alternatives} \label{Table}
\vspace{0.5cm}
\begin{tabular}{>{\raggedright\arraybackslash}p{0.3\textwidth} >{\raggedright\arraybackslash}p{0.35\textwidth} >{\raggedright\arraybackslash}p{0.3\textwidth}}
\toprule
\textbf{Platform-controlled AI agents} &
\textbf{Hardware-ecosystem agents} &
\textbf{User-centric agent advocates} \\
\midrule
\textbf{Hypothetical Example: }
Amazon Alexa Pro---a platform agent that collects all user interactions, prioritizes Amazon services and sponsored products, manipulates user attention through notifications, and creates significant switching costs to leave the Amazon ecosystem. &
\textbf{Hypothetical Example: }
Banana Intelligence Plus---an agent operating primarily within hardware company Banana's hardware ecosystem, emphasizes privacy while collecting data for model improvement, offers fair third-party options while subtly prioritizing Banana's other services, and primarily attempts to reinforce hardware purchases. &
\textbf{Hypothetical Example: }
Open Assistant---a fully open-source agent that runs completely locally, stores all data on user devices, interacts with any market or service without bias, allows full user control over filtering and priorities, and can be easily migrated between devices or platforms. \\ 
\midrule
\textbf{Lock-in via platform agents. }
Platform agents become indispensable ``digital concierges,''
erecting switching barriers and concentrating control. &
\textbf{Lock-in to a hardware ecosystem. }
Agents are tied to specific hardware products but may offer limited interoperability within the ecosystem. &
\textbf{Fully interoperable. }
Agent advocates can run on local or diverse hardware systems,
making it easier to switch or migrate data and preserving user choice. \\ 
\midrule
\textbf{Invasive personal data collection. }
Platform agents observe every digital interaction, enabling
unprecedented user profiling and heightened surveillance. &
\textbf{Data collection not used for surveillance. }
Collects data primarily for improving models and services, with some privacy protections in place. &
\textbf{Local data.}
Agent advocates enable users to store and process data locally,
minimizing external data access and surveillance risks. \\ 
\midrule
\textbf{Repressive market design. }
Platform agents can steer transactions to certain vendors,
embed hidden fees, and block competitor access. &
\textbf{Transparent arbitrator with fair terms. }
Attempts to provide fair options but may have subtle preferences for ecosystem-compatible services. &
\textbf{Open market interactions. }
Agent advocates can browse, compare,
and transact across diverse vendors, thereby reducing platform dependence. \\ 
\midrule
\textbf{Manipulating attention and engagement. }
Platform agents prioritize content and
notifications aligned with platform interests. &
\textbf{Interested in complementary services, not in market manipulation. }
Focuses on enhancing hardware value rather than driving engagement metrics. &
\textbf{Enhanced user autonomy. }
Agent advocates enable users to define how content is
filtered or ranked, reducing platform influence over content consumption. \\
\bottomrule
\end{tabular}
\caption{There is a spectrum from platform-controlled agents to fully user-centric agent advocates. We illustrate both ends of the spectrum as well as an approximate midpoint.}
\end{table*}

\end{document}